\documentclass{desyproc}

\begin{document}
\title{The Total Cross Section at the LHC:\\
Models and Experimental Consequences}

\author{{\slshape J.R. Cudell }\\[1ex]
IFPA, AGO Dept., Universit\'e de Li\`ege, Belgium}

\contribID{cudell\_jr\_total}

\desyproc{DESY-PROC-2009-xx}
\acronym{EDS'09} 
\doi  

\maketitle

\begin{abstract}
I review the predictions of the total cross section for many models, and point
out that some of them lead to the conclusion that the standard experimental analysis
may lead to systematic errors much larger than expected.
\end{abstract}

The total cross section is  a highly non-perturbative object that we cannot predict from QCD. In fact,
we have to rely on theoretical ideas that were developed before QCD, in the context of the analytic 
$S$ matrix, such as analyticity, or the unitarity of partial waves.  

The natural place to discuss these ideas is the complex-$j$ plane, where the singularities of the 
amplitudes determine their behaviour with $s$. But we do not know these singularities. The simplest 
ones, which correspond to the exchange of bound states, are simple poles. Accounting for the exchange of 
meson trajectories is not too hard, as we know their spectrum. But as their contribution falls with $s$, 
they will not matter at the LHC. One must thus model the pomeron, for which there is little 
spectroscopic guidance. Again, the simplest idea is to use a simple pole at $j=1+\epsilon$. But we know 
that if there are simple poles, there must be cuts, which correspond to multiple exchanges. We do not 
know how to calculate these: we only know general properties of the two-pomeron cut. This means that 
there 
are many possibilities, such as eikonal models, $U$-matrix unitarisation, extended eikonal/$U$-matrix 
models, or multi-channel eikonals. These cuts will be needed at the LHC to restore partial-wave 
unitarity. It is also possible that the pomeron is not an exchange of bound states, so that one should 
not start from a simple pole, but rather consider multiple poles (double or triple) at $j=1$, which 
automatically obey unitarity.
 
\section{The COMPETE fits}
The multitude of models has been confronted with data, and the COMPETE collaboration~\cite{COMPETE1} 
gathered and cleaned 
the datasets which are now available on the servers of the Particle data group~\cite{PDG}, and which 
gather all
the soft data measured at $\sqrt{s}\geq 4$ GeV and $t=0$, {\it i.e.} 
the total cross sections and the $\rho$ 
parameters when 
available, for $pp$, $\bar p  p$, $\pi^\pm p$, $K^\pm p$, $\gamma p$ and $\gamma \gamma$. These data 
have some drawbacks. First of all, most cross sections have been measured at ISR energies, and there
is a gap from $\sqrt{s}\approx 100$ GeV to $\sqrt{s}\approx 500$ GeV. This gap would have been filled by 
the pp2pp collaboration at RHIC, but it has unfortunately been stopped. Another problem comes from the 
highest-energy data, where a $2\sigma$ disagreement exists between E710 and CDF. Finally, the 
compatibility of the various measurements of $\rho$ at ISR energies is questionable, and not enough
information has been published to redo the analyses.

\begin{wrapfigure}{r}{0.45\textwidth}
\centerline{\includegraphics[width=0.4\textwidth]{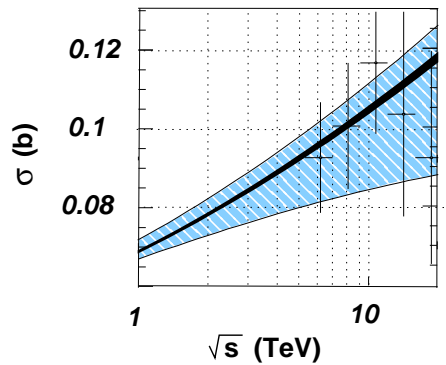}}
\caption{The COMPETE prediction for the total cross section in the LHC energy region. The inner black 
band represents our best guess with statistical errors only, the outer band all the parametrisations with $\chi^2$/point $\leq 1$.}
\label{Fig:compete}
\end{wrapfigure}

Nevertheless, COMPETE considered several hundreds of possible parametrisations, based on simple, double 
or triple poles, and kept only those which had a global $\chi^2$/point smaller than 1, for 
$\sqrt{s}\geq 5$ GeV. From these, one can predict the total cross section at the LHC, and
estimate the error due to the extrapolation. Figure 1 shows the result of this study. 
The extrapolation of the parametrisations 
which passed the $\chi^2$ test is given by the outer band. It predicts 84 mb $\leq\sigma\leq 112$ mb
for $\sqrt{s}=10$~TeV, and 90--117 mb for 14 TeV. 

The inner band of Fig.~1 shows the best guess, and it is based 
on a series of statistical indicators explained in~\cite{indicators}. It corresponds to a universal triple-pole ($\log^2 s$) parametrisation. 
One should note that the multiple-pole fits have some peculiar properties: in the triple-pole case, the 
contribution of the pomeron $falls$ with $s$ for energies smaller than 6 GeV, whereas in the 
double-pole case, it actually becomes negative below 9.5 GeV. Simple poles seem to be excluded, but they 
can be considered again if the lower cut-off on the energy is raised to 10 GeV.
The multiple-pole parametrisations have been used successfully~\cite{cudellsoyez} to reproduce
low-$x$ data in deeply inelastic scattering (DIS), showing that universality does
not seem to hold in general, as the coefficient of the triple pole depends on $Q^2$. More recently,
multiple poles have been used successfully~\cite{genya} to reproduce the differential elastic cross 
sections~\cite{elasticdataset}. 

Hence multiple-pole fits seem to work, but it is hard to imagine how such a structure may emerge from 
QCD. On the other hand, simple-pole fits have a simple phenomenology, a simple interpretation as the
exchange of glueballs, and only a handful of parameters. 

\section{The simple-pole fit: low energy}
Clearly, one simple-pole pomeron has never been a reasonable choice: in their original 
soft-pomeron model~\cite{DoLa1}, Donnachie
and Landshoff  included a two-pomeron cut, which lowers the cross section, and 
which enables one to go to $\approx 40$ TeV before unitarity is broken. However, such a simple
model seems to produce worse fits than multiple poles. Furthermore, HERA data on DIS and 
on vector-meson production have lead to the introduction of a second singularity 
\cite{DoLa2}, taken also as a simple pole, and mimicking the BFKL cut. One can indeed follow the 
standard transformation to the $j$-plane to convince oneself that hadronic singularities must be
present in DIS~\cite{CuDoLa}. This new singularity has a coupling which
depends on $Q^2$, or on the vector-meson mass, and, together with the soft pomeron, it enables
one to reproduce all HERA measurements~\cite{DoLa3}. The question then is to know whether such a 
singularity might be present in soft data.

\begin{wrapfigure}{r}{0.45\textwidth}
\centerline{\includegraphics[width=0.4\textwidth]{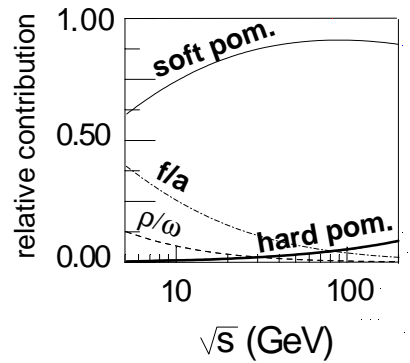}}
\caption{The contribution to the total $pp$ cross section of
 the two pomerons and the two reggeons, divided by the average of the $pp$ and $\bar p p$ total cross 
 sections.}
\label{Fig:relative}
\end{wrapfigure}

It actually came as a surprise that, indeed, the introduction of a hard pole brings the 
simple-pole pomeron fit on equal footing with multiple poles, as far as the $\chi^2$/point is 
concerned~\cite{CLMS}. This is true provided that one stops the fit at $\sqrt{s}\approx 200$ GeV. Beyond 
that, the hard pole must be unitarised, as its contribution to the total cross section would be much 
too large. Nevertheless, it is remarkable that a global fit to soft data leads to a hard pomeron with an 
intercept $\approx 1.4$, perfectly compatible with that obtained by Donnachie and Landshoff in DIS. It 
must be noted however that the coupling of the hard pomeron is tiny. It contributes at most 7\% to the 
total cross section for $\sqrt{s}<200$~GeV, as shown in Fig.~\ref{Fig:relative}. Previous 
studies~\cite{COMPETE1} considered only fits to the highest energy, in which case the hard pomeron 
coupling becomes extremely small. Even if the energy is limited to 200 GeV, the coupling remains of the 
order of 1\% of that of the soft pomeron.

The two-pomeron model was later applied to the differential elastic cross section in~\cite{CLM},
where it was shown that the hard pomeron was compatible with data on the first cone (0.1 GeV$^2<|t|\leq 
0.5$ GeV$^2$), for energies smaller than 100 GeV, although in this case it did not significantly 
improve the fit. Note also that the form factors of the various exchanges were directly extracted from
the data, confirming the presence of a zero in the form factor of the $C=-1$ trajectory,
and suggesting one for the $c=+1$ meson trajectory. 

One has thus found an alternative to the multiple-pole fits: the two-pomeron model, which can reproduce 
DIS data, 
photoproduction data, soft forward data and differential elastic cross section, provided one stays
at moderate energy and small $|t|$. As we shall now see, if this model is the correct one, then it
may have important consequences at the LHC.

\section{The simple-pole fit: high energy}

Indeed, to go to higher energies, one must confront the question of unitarisation. At high energy, the
partial waves can be obtained by a Fourier transform of the amplitude ${\cal A}(s,t)$ to 
impact-parameter 
space, as $\ell\approx b\sqrt{s}$. The unitarity constraint can then be written 
\begin{equation}
\left|{\cal A}(s,b)\right|^2\leq 2 Im\left({\cal A}(s,b)\right).\nonumber
\end{equation}
It is easy to see that this means that the complex ${\cal A}(s,b)$ has to be in a circle of radius 1
centred at ${\cal A}(s,b)=i/2$, which we shall call the unitarity circle (Note that another 
normalisation is frequently used, where the circle has radius 1 and is centred at $i$.)

\begin{figure}[ht]
~\vglue 0.3cm
\begin{minipage}[b]{0.45\linewidth}
\centering
\includegraphics[width=0.8\textwidth]{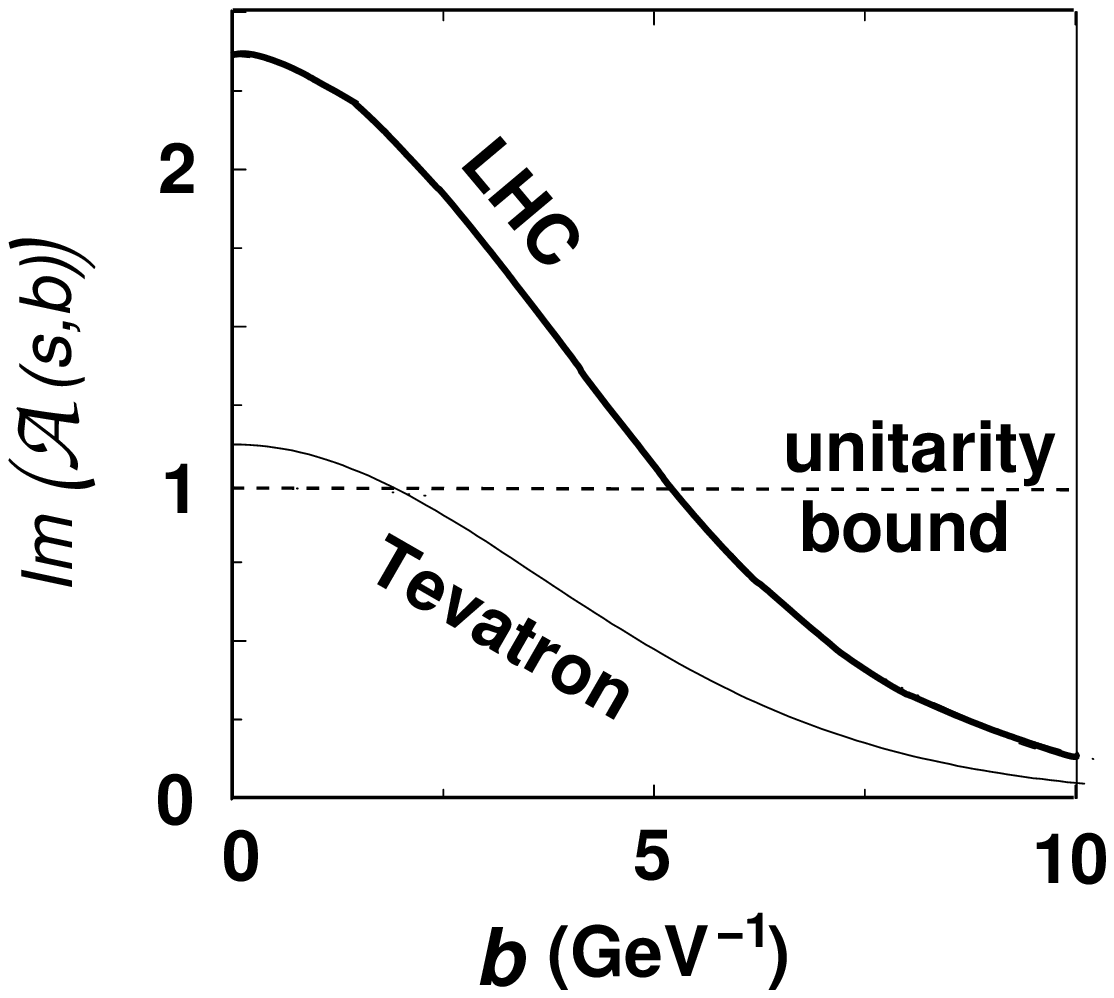}\vglue -0.5cm
\caption{The imaginary part of the amplitude ${\cal A}(s,b)$ at the Tevatron and at the LHC, before unitarisation.}
\label{Fig:unita}
\end{minipage}
\hspace{1cm}
\begin{minipage}[b]{0.45\linewidth}
\centering
\includegraphics[width=0.8\textwidth]{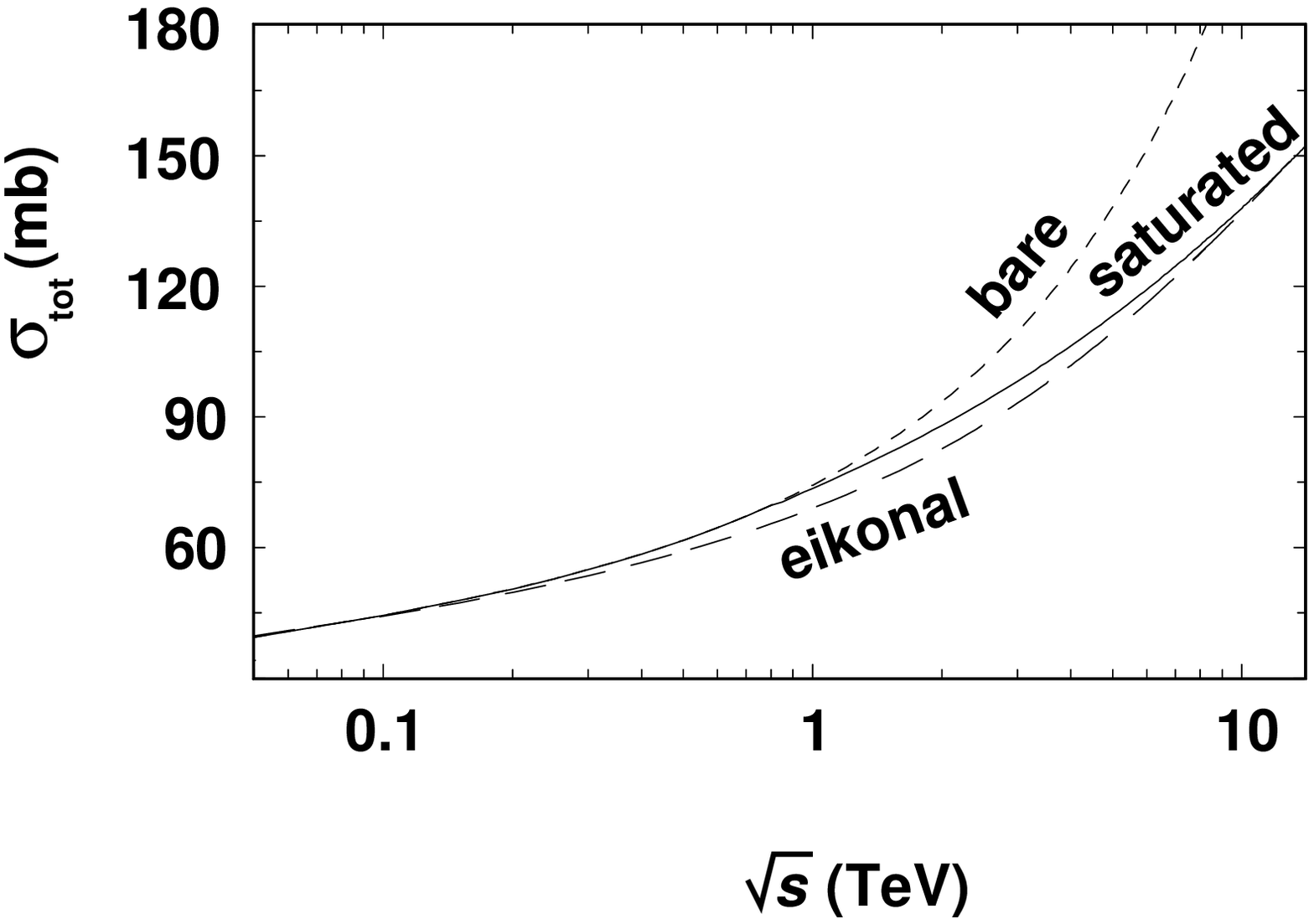}
\caption{Total cross section in a two-pomeron model without (bare) and with simple unitarisation 
schemes.}
\label{Fig:cross}
\end{minipage}\end{figure}

In our two-pomeron model, the amplitude ${\cal A}(s,0)$ leaves the unitarity circle for energies
slightly below the Tevatron, and it is quite far out at the LHC, as can be seen in Fig.~\ref{Fig:unita}.
The problem, of course, is that one does not know exactly how the amplitude will stay in the unitarity 
circle. It is known that multi-pomeron exchanges, and maybe multi-pomeron vertices, will restore 
unitarity. But how to implement these ingredients is far from unique~\cite{PVL2}. Even the simplest 
two-pomeron cut is not fully calculable, let alone a full resummation.

Mathematically, unitarisation schemes can be classified into three classes. The first one maps the whole 
upper-half complex plane onto the unitarity circle. In this class, one finds the eikonal scheme, and the
$U$-matrix scheme~\cite{Umat}. Note that the latter is the only one to provide a one-to-one mapping 
between the
half-plane and the circle, but the amplitude also has a new pole at ${\cal A}(s,b)=-2i$. 
The second class is a simple extension of the first one: instead of mapping
the upper half-plane to the full unitarity circle, one maps it to a smaller circle, still centred
at ${\cal A}(s,b)=i/2$. The third class, which we recently proposed~\cite{CPS}, maps a given amplitude
to the circle, which is far less restrictive than mapping the half plane. In general, these schemes can 
have very different properties~\cite{CPS}: they can lead to the Pumplin bound
($\sigma_{el}<\sigma_{tot}/2$~\cite{Pump}) and to shadowing, but it is also possible to have 
anti-shadowing and $\sigma_{el}\approx\sigma_{tot}$, both for eikonal-like and $U$-matrix-like schemes.
 I shall not consider these more exotic
possibilities here, but one has to bear in mind that they cannot be a priori ruled out.

As we do not know what scheme to use, we have decided to be 
conservative, and to unitarise the two-pomeron 
model using the simplest and most conservative approach. The minimal scheme assumes that something
happens when the amplitude reaches the unitarity bound, and one simply cuts-off the amplitude at 1.
We shall call it the saturation scheme.
Note that to do that while preserving analytic properties is not very simple, but nevertheless
it can be done~\cite{CS1}. The other scheme is the standard one-channel eikonal.

Using either of these schemes produces an amplitude which respects unitarity, and thus can be used to
predict the total cross section at the LHC. The first scheme keeps by definition the results of the fit 
below the Tevatron energy, and thus one can keep the parameters of the low-energy simple-pole fit. The 
eikonal on the other hand changes slightly the low-energy results, and thus necessitates some re-
fitting. The result is shown in Fig.~\ref{Fig:cross}. One sees that, in this very simple approach, the
total cross section could be as large as 150 mb. It is interesting to note that a similar number
has been obtained by Landshoff, in a very different manner~\cite{PVL2}.

\section{The measurement of $\sigma_{tot}$}

\begin{wrapfigure}{l}{0.45\textwidth}
\centerline{\includegraphics[width=0.4\textwidth]{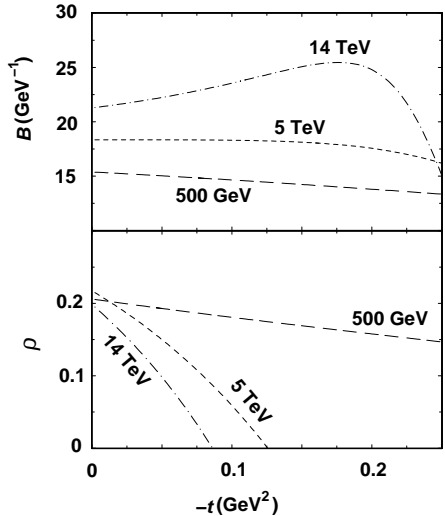}}
\caption{The slope $B$ and the ration of the real  to the imaginary parts of the amplitude for $pp$ scattering, for various energies, in the saturation scheme.}
\label{Fig:rhoB}
\end{wrapfigure}

The unitarised two-pomeron model has a consequence that has been overlooked so far. As the energy grows,
the protons become blacker, and edgier. This in turn changes the diffraction pattern, {\it i.e.} the
exponential slope of the cross section varies quickly with $t$: the differential cross section
cannot be approximated any longer by $e^{Bt}$ with $B$ constant. In turn, the real part of the amplitude 
also develops a strong $t$ dependence, and $\rho(0)$ could be as large as 0.24. We have checked that 
this is the case for the eikonal scheme, and for
the saturation scheme~\cite{CS1}, and show the results for the latter in Fig.~\ref{Fig:rhoB}.

A hadronic slope $B$ almost constant with $t$ for $|t|<0.25$ GeV$^2$ and 
a small $\rho$ parameter are standard 
assumptions for the planned measurement of the total cross section~\cite{TOTEM}. If the two-pomeron
model is correct, then these assumptions will be wrong, and we can evaluate the systematic 
uncertainty that they will generate. 
So, we have used our unitarised two-pomeron model~\cite{CS2} to simulate data, 
using bins and errors similar to those of the UA4/2 
experiment~\cite{UA42}, and including the Coulomb-nucleon interference~\cite{CNI}. We then performed
the standard analysis on these simulated data, and extracted a measurement of the total cross section,
which we could compare to the input value from our model. The result of this analysis is that the 
extracted value of the total cross section will systematically overshoot the model value by about 10 mb
 in a luminosity-dependent method, and about 15 mb in a luminosity-independent one.
Hence, an additional study of the $t$ dependence of the
slope and of the $\rho$ parameter will be needed before one reaches the 1 \% precision level. 
\\

\section{Conclusion}
The measurement of the total cross section at the LHC will tell us a lot about the analytic structure of
the amplitude, as there is a variety of predictions that span the region from 90 to 230 mb:
\begin{itemize}
\item $\sigma_{tot}>200$ mb: the only unitarisation scheme able to accommodate such a large number is
the $U$ matrix~\cite{Umat}. It basically predicts the same inelastic cross section as more
standard schemes, but the elastic cross section is much larger, and accounts for the difference.
\item 120 mb $<\sigma_{tot}< 160$ mb: this would be a clear signal for a two-pomeron model, and would also
tell us about the unitarisation scheme.
\item $\sigma_{tot}\approx 110$ mb: this is the standard prediction not only of the COMPETE fits, but
also of many models based on a simple eikonal and only one pomeron pole.
\item $\sigma_{tot}<100$ mb: this would indicate either the validity of double-pole parametrisations, or 
that of unitarisation schemes in which multiple-pomeron vertices are important~\cite{Tel,Dur}.
\end{itemize}

And finally, if one really enters a new regime at the LHC, the study of the $t$ dependence 
of the elastic cross section, and of the real part of the amplitude, may be of great importance
to unravel the underlying dynamics and to improve the experimental measurement of $\sigma_{tot}$.

\section*{Acknowledgements}

I acknowledge the contribution of my collaborators of the COMPETE collaboration 
to the investigations summarised here,
and in particular of A.~Lengyel, E.~Martynov and O.V.~Selyugin. I also acknowledge discussions
with A. Donnachie, P.V. Landshoff, U. Maor and S. Troshin, and thank Karine Gilson for a careful
proofreading.

\begin{footnotesize}

\end{footnotesize}
\end{document}